\documentclass[aps,prb]{revtex4}
\usepackage{graphicx,amsbsy,bm}

\begin{document}

\title{Radius dependent shift of surface plasmon frequency in large metallic nanospheres: theory and experiment}
 \author{W. Jacak$^1$, J. Krasnyj$^{1,2}$, J. Jacak$^1$, R. Gonczarek$^1$, A. Chepok$^2$, L. Jacak$^1$, D. Z. Hu$^3$, and D. Schaadt$^3$}

\affiliation{$^1$Institute of Physics, Wroc{\l}aw University of Technology,
Wyb. Wyspia\'{n}skiego 27, 50-370, Wroc{\l}aw, Poland;
$^2$ Theor. Phys. Group, International  University, Fontanskaya Doroga 33, Odessa, Ukraine;
$^3$ Institute of Applied Physics/DFG-Center for Functional Nanostructures, Karlsruhe Institute of Technology, Karlsruhe, Germany}

\begin{abstract}
Theoretical description of oscillations of  electron liquid in large metallic nanospheres (with radius of few tens nm)
 is formulated within random-phase-approximation semiclassical scheme. Spectrum of plasmons
 is determined including both  surface and volume type excitations. It
   is demonstrated that only surface plasmons of dipole type can be excited by homogeneous dynamical electric field.
   The Lorentz friction due to irradiation of electro-magnetic wave by plasmon oscillations  is analyzed
    with respect to the sphere dimension. The resulting shift of resonance frequency   turns out to be strongly sensitive
    to the sphere radius. The form of e-m response of the system of metallic nanospheres embedded in
     the dielectric medium is found. The theoretical predictions are verified by a measurement of extinction of light
      due to plasmon excitations in nanosphere colloidal water solutions, for Au and Ag    metallic components with radius
       from 10 to 75 nm. Theoretical predictions and experiments clearly agree in the positions of surface plasmon resonances and  in
        an emergence of the first volume plasmon resonance in the e-m response of the system for limiting big nanosphere radii, when dipole
        approximation is not exact.
\end{abstract}

\vspace{0.1mm}
\vspace{0.1mm}

\maketitle

\section{Introduction}
Experimental and theoretical investigations of plasmon excitations in metallic nanocrystals have received recently
much attention due to possible applications in photo-voltaics and microelectronics. A significant enhancement of absorption
 of the incident light in photodiode-systems with active surface covered with  metallic particles (of Au, Ag or Cu), with radius
   ten to several tens nanometers and
  with planar density $\sim 10^8$/cm$^2$, was observed\cite{wzmocn1,wzmocn2,wzr1,wzr2,wzr3,wzr4,wzr5}. This is due to mediating of
  light energy transfer by surface plasmon oscillations in metallic nano-components. These findings are of
   practical importance towards enhancement of solar cell efficiency especially for thin film cell technology. On the other side, hybridized
   states of  surface plasmons and photons result in plasmon-polaritons\cite{maradudin} which are of high significance for
   applications in sub-diffractional photonics and microelectronics\cite{zastos,plasmons}

For finite size crystals where the surface strongly affects the plasmon spectrum, surface plasmons occur and dominate the electro-magnetic
(e-m) response of the metallic  system. A strong dependence of resonance surface plasmon frequencies on the nanoparticle
size and shape are reported in the literature\cite{wzr2,burda}.
The plasmon resonances of noble metals, such as gold and silver (including also copper), are of particular
interest  due to their frequencies located within the visible part of the e-m spectrum.

Plasmon oscillations in metallic nanospheres can be excited  by time-dependent electric field signal.
There are different types of plasmon oscillations in the case
  of metallic nanoparticle. General types are volume and surface plasmons, referring to oscillations of internal electron density
  and surface electron density, respectively.  The surface plasmons are linked to
  translational motion of all electrons which results  in surface density oscillations only. The volume modes are related to
   compressional oscillations. Note that separation
  of these both types of collective excitations of electrons  in metallic
   nanocrystal repeats the similar distinguishing of collective modes in atomic nuclei\cite{migdal,steinw,goldb} (confirmed
   by giant resonance experiments).
  For metallic ultra-small clusters with  number $N$ of electrons  (in the range form $N= 8$ to $N=200$ in  Na clusters)\cite{brack} the decoupling
  of volume and surface excitations is demonstrated by microscopic modeling\cite{brack} at approximately $N=50$. For ultra-small metallic clusters quantum shell
  effects\cite{brack} and spill-out of electron cloud beyond the ionic jellium\cite{ekardt,weick} disturb separate
   formation of surface and volume collective excitations, while for large clusters (with radius larger than 10 nm) the role of shells and spill-out is considerably
   reduced\cite{brack,ekardt,kresin}, and both modes are well defined.

  In order to excite the volume type oscillations an electric dynamical field inhomogeneous on the  nanosphere scale  is necessary while a homogeneous field excites only surface plasmons.
   For spherical symmetry, all modes of plasmon oscillations
can be represented by spherical harmonics in terms of $l,m$, angular momentum (multiplicity) numbers.
A  dynamical electric field homogeneous over the sphere can induce only $l=1$, i.e., dipole-type surface oscillations.

The surface plasmons have been originally considered by Mie\cite{mie}, who provided a classical description of oscillations of electrical
 charge on the surface of the metallic sphere within the classical model. The dipole-type Mie oscillation energy is
  not dependent of the sphere radius, in contradiction to experimental observations, both in the case of small and larger nanospheres.
For low radius, of order of single nanometers, besides mentioned above spill-out, the electron-electron interactions are
 important\cite{burda,brack,ekardt,kresin} (including decay of plasmons into particle-hole pairs with similar energy,
 called as Landau damping\cite{weick,weick1})---these quantum effects influence on position of resonance frequency.
For large spheres, with radius bigger than 10~nm, even stronger shifts of resonance are observed, probably
connected  with another mechanism, since with radius growth
quantum effects turn out to be  not so important as for ultra-small clusters.
The  case of bigger nanospheres is, however,  of particular significance as such  metallic nano-components would be applied to
 enhancement of photo-voltaic effect in metallically modified solar cells\cite{wzmocn1,wzmocn2,wzr1,wzr2,wzr3,wzr4}.

Plasma excitations in metallic clusters were analyzed within many attitudes,  addressed, however, mostly to small clusters.
In particular were developed  numerical methods of  calculus 'ab initio' including  Kohn-Sham 'local density approximation [LDA], similar
as applied in chemistry for large molecule calculations (limited, however, to few hundreds of electrons)\cite{serra,ekardt,brack}.
Also  variational methods for energy density, semiclassical approach\cite{kresin} and  random-phase-approximation (RPA) numerical summations were applied
 (e.g., for clusters of Na with radius $\sim 1$nm)\cite{brack1}. Emerging of the Mie response from the more
  general description was analyzed, but including only (in a numerical manner) single breathing  volume mode\cite{brack1}.
Commonly the 'jellium' model was applied, allowing for adiabatic approach to background ion system. In the 'jellium' model all the kinetics concerns
electron liquid screened by uniform and static  background of positive ions\cite{brack,ekart,ksi}.

Below we present the simplified RPA-type theory of plasmon excitations in metallic nanosphere embedded in dielectric medium adjusted to large
  nanospheres (with radius above 10 nm),
when quantum corrections beyond semiclassical approximation are not so significant as in the case of ultra-small clusters\cite{brack,ekardt,kresin}.
 For  optically induced plasma
  oscillations, the wavelength of incident light which excites resonance oscillations in metallic nanospheres (Au and Ag within several to several tens
   nm for  radius) is considerably longer ($\lambda\sim 400$ nm) than nanosphere dimension. Thus, the dipole-type approximation is
    valid, i.e., one can  assume that the electric field of the incident e-m wave is homogeneous over the nanosphere. Therefore, only dipole type
     oscillations of  surface plasmons  will contribute the resonance (except for limiting big nanospheres, when also volume excitations seem
     to enter   e-m response due to not exact dipole approximation).

The  experimentally observed red-shift of dipole Mie  plasmon resonance for ultra-small clusters is caused mainly by significant  spill-out effect reducing density
of electrons\cite{brack,ekardt,weick,kresin}. The Mie frequency is proportional to square root of the electron density and thus one arrives with
reduced its value due to spill-out beyond the edge of ion jellium.
With
growing radius this effect weakens, as being  of surface type and thus proportional to inverse radius,  quite oppositely as red-shift observed for bigger
 nanospheres. For the radius
 over 10 nm the red-shift experimentally observed is much stronger than that one  for small clusters and is sharply growing with radius enhancement.
In order to explain this phenomenon observed\cite{wzr2}  in nanospheres of Au and confirmed by more precise measurements for Au and Ag nanospheres
reported in the present paper,
we have included damping of plasmon oscillations
via irradiation effects, which seem to dominate plasmon energy losses at larger scale of radii and  cause strong red-shift of the resonance.
 We performed  measurements of light extinction by nanospheres (in water colloidal solution)
 of Au with radii from 10 nm to 75 nm and Ag from 10 nm to 40 nm.
  The resulting data reveal a strong shift of
resonance towards higher wave-lengths with  radius growth. Radiation losses, which, as we suppose,  are responsible for radius dependent
red-shift of resonance frequency, can be  described in terms of Lorentz friction\cite{lan}, and calculated also independently in Poyting vector
 terms in a far-field zone of plasmon radiation. Damping of plasmons is caused also additionally by
  electron scattering processes and we verify that this channel is not important for $a>20$ nm ($a$---nanosphere radius)
   when radiation losses are much stronger. The shift of the resonance frequency of dipole-type surface plasmons
    resulting due to damping phenomena is compared with the experimental data for various nanosphere radii. Emerging of the first mode of volume plasmons
     is experimentally observed for $2a\sim 150$ nm for Au and $2a\sim 80$ nm for Ag, due to breaking the dipole approximation at this range
     of radius, in agreement with the presented theory predictions.

The paper is organized as follows. In the first paragraph, the RPA theory\cite{rpa,pines,jac} is
 generalized for the confined system of spherical shape. In the second one, the equations
  for volume and surface plasmons are solved (with particularities of calculus in the Appendix). The third paragraph contains
   a description of the Lorentz friction for surface plasmons oscillations of the dipole-type. In the fourth one, an analysis
    of the radiation losses is presented using the Poyting vector of plasmon radiation in far-field region, which supports the
     previous Lorentz friction estimation. The following paragraph comprises the summarizing of whole e-m response of metallic nanosphere system.
     The  last paragraph presents a comparison of the theoretical predictions with
      experimental data of  e-m response features of the colloidal water solutions with nanospheres of Au and Ag with several radii of metallic nano-components.

\section{RPA semiclassical  approach  to electron excitations in metallic nanosphere}
\subsection{Derivation of RPA equation for local electron density in spherical geometry}
Let us consider a metallic sphere with a radius $a$, located for the starting model in the vacuum, $\varepsilon = 1,\;\mu =1$
 and in the presence of a dynamical electric field (magnetic field is assumed zero). The model {\it jellium}\cite{brack,ekart,ksi}
  is assumed in order to account for the screening background of positive ions  in the form of static uniformly distributed over the sphere positive charge:
\begin{equation}
n_e({\bm r})=n_e \Theta(a-r),
\end{equation}
where  $n_e=N_e/V$ with $n_e|e|$ is the averaged positive charge density, $N_e$ is the number of collective electrons in the sphere, $V=\frac{4 \pi a^3}{3}$ is
the sphere volume,
and $\Theta$ is the Heaviside step-function. After neglecting  the ion dynamics within the {\it jellium} model,
 which can be  adopted in particular for description of simple metals, as noble, transition and alkali metals, we deal with the Hamiltonian for collective electrons,
\begin{equation}
\label{h1}
 \hat{H}_{e} = \sum\limits_{j =1}^{N_e} \left[ -\frac{\hbar^2 \nabla_{j}^2}{2m} -e^2 \int\frac{n_e ({\bm r_0})d^3 {\bm r_0}}{|{\bm r}_{j}
   -  {\bm r_0}|} +e\varphi({\bm r}_j, t)\right]
 +\frac{1}{2}\sum\limits_{j\neq j'}\frac{e^2}{|{\bm r}_{j}-{\bm r}_{j'}|} +\Delta E,
\end{equation}
where ${\bm r}_{j}$ and $m$ are the position (with respect to the dot center) and mass of $j$th electron, $\Delta E$
 represents the electrostatic energy contribution from the ion jellium, and $\varphi({\bm r}, t)$ is the scalar potential of the external electric field.
 The corresponding electric field ${\bm E}({\bm r}_j, t)=-\nabla\varphi({\bm r}_j, t)$. Assuming that space-dependence of ${\bm E}$ is weak on the scale of the sphere radius $a$ (i.e the electric field is homogeneous over the sphere), $\varphi({\bm r}_j, t)=-{\bm r}_j\cdot\bm E(t)$.

A local electron density can be written as follows\cite{rpa}:
\begin{equation}
\rho({\bm r}, t)=<\Psi({\bm r_e},t)|\sum\limits_j \delta({\bm r}-{\bm r}_j) |\Psi({\bm r_e},t)>,
\end{equation}
$ i\hbar \frac{\partial\Psi({\bm r_e},t)}{\partial t}=\hat{H}_{e}\Psi({\bm r_e},t)$,  ${\bm r_e}=({\bm r_1},{\bm r_2},...,{\bm r_N})$, with the Fourier picture:
\begin{equation}
\tilde{\rho}({\bm k}, t)=\int \rho({\bm r},t) e^{-i{\bm k}\cdot {\bm r}} d^3 r = <\Psi'({\bm r_e},t)|\hat{\rho}({\bm k})|\Psi'({\bm r_e},t)>,
\end{equation}
where the 'operator' $ \hat{\rho} ({\bm k})=\sum\limits_{j}  e^{-i{\bm k}\cdot {\bm r_j}} $.

Using the above notation one can rewrite $\hat{H}_{e}$, in analogy to the bulk case\cite{pines}, in the following form:
\begin{equation}
\begin{array}{l}
\hat{H}_{e} = \sum\limits_{j =1}^{N_e} \left[ -\frac{\hbar^2 \nabla_{j}^2}{2m}\right]
-\frac{e^2}{4 \pi^2} \int d^3 k \tilde{n}_e({\bm k}) \frac{1}{k^2} \left(\hat{\rho^{+}}({\bm k}) +  \hat{\rho}({\bm k})\right) \\
 +\frac{e^2}{16 \pi^3} \int d^3 k \tilde{\varphi}({\bm k},t) \left(\hat{\rho^{+}}({\bm k})
  +  \hat{\rho}({\bm k})\right) +\frac{e^2}{4 \pi^2}\int d^3 k
  \frac{1}{k^2}\left[ \hat{\rho^{+}}({\bm k})  \hat{\rho}({\bm k}) -N_e \right]
  +\Delta E,\\
  \end{array}
\end{equation}
where:
$\tilde{n}_e({\bm k})=\int d^3 r n_e ({\bm r}) e^{-i{\bm k}\cdot {\bm r}}$,
$\frac{4 \pi}{k^2}= \int d^3 r \frac{1}{r} e^{-i{\bm k}\cdot {\bm r}}$,
$\tilde{\varphi}({\bm k})=\int d^3 r \varphi({\bm r},t) e^{-i{\bm k}\cdot {\bm r}}$.

Utilizing this form of the electron Hamiltonian one can write the motion equation for $\hat{\rho} ({\bm k})$:
\begin{equation}
\frac{d^2 \hat{\rho} ({\bm k}) }{dt^2}=\frac{1}{(i\hbar)^2} \left[\left[ \hat{\rho} ({\bm k}),\hat{H}_{e} \right],\hat{H}_{e} \right],
\end{equation}
or, after some algebra:
\begin{equation}
\begin{array}{l}
\frac{d^2 \delta \hat{\rho} ({\bm k}) }{dt^2}=-\sum\limits_{j}e^{-i{\bm k}\cdot {\bm r}_j}\left\{ -\frac{\hbar^2}{m^2}\left(
{\bm k}\cdot \nabla_j \right)^2
+ \frac{\hbar^2 k^2}{m^2}i  {\bm k}\cdot \nabla_j +\frac{\hbar^2 k^4}{4 m^2}\right\}\\
-\frac{ e^2}{m 2\pi^2}\int d^3q  \tilde{n}_e ({\bm k}-{\bm q})\frac{{\bm k}\cdot {\bm q}}{q^2} \delta \hat{\rho}( {\bm q})
-\frac{ e}{m 8\pi^3}\int d^3q  \tilde{n}_e ({\bm k}-{\bm q})({\bm k}\cdot {\bm q})\tilde{\varphi}({\bm q},t)\\
-\frac{ e}{m 8\pi^3}\int d^3q  \delta\hat{\rho} ({\bm k}-{\bm q})({\bm k}\cdot {\bm q})\tilde{\varphi}({\bm q},t)
-\frac{e^2}{m 2 \pi^2}\int d^3q \delta \hat{\rho}({\bm k}- {\bm q})\frac{{\bm k}\cdot {\bm q}}{q^2}  \delta \hat{\rho}( {\bm q}),\\
\end{array}
\end{equation}
where $\delta \hat{\rho}({\bm k}) =  \hat{\rho}({\bm k)} -  \tilde{n}_e ({\bm k})$ is the 'operator' of local electron density fluctuations beyond the uniform distribution.
Taking into account that:
$\delta \tilde{\rho}({\bm k},t)=   <\Psi(t)|\delta\hat{\rho}({\bm k})|\Psi(t)>=  \tilde{\rho}({\bm k},t) -  \tilde{n}_e ({\bm k})$
we find:
\begin{equation}
\label{e10}
\begin{array}{l}
\frac{\partial^2 \delta \tilde{\rho} ({\bm k},t) }{\partial t^2}=<\Psi|-\sum\limits_{j}e^{-i{\bm k}\cdot {\bm r}_j}\left\{ -\frac{\hbar^2}{m^2}\left(
{\bm k}\cdot \nabla_j \right)^2
+ \frac{\hbar^2 k^2}{m^2}i  {\bm k}\cdot \nabla_j +\frac{\hbar^2 k^4}{4 m^2}\right\}|\Psi>\\
-\frac{ e^2}{m 2\pi^2}\int d^3q  \tilde{n}_e ({\bm k}-{\bm q})\frac{{\bm k}\cdot {\bm q}}{q^2} \delta \tilde{\rho}( {\bm q},t)
-\frac{ e}{m 8\pi^3}\int d^3q  \tilde{n}_e ({\bm k}-{\bm q})({\bm k}\cdot {\bm q})\tilde{\varphi}({\bm q},t)\\
-\frac{ e}{m 8\pi^3}\int d^3q  \delta\tilde{\rho} ({\bm k}-{\bm q},t)({\bm k}\cdot {\bm q})\tilde{\varphi}({\bm q},t)
-\frac{e^2}{m 2 \pi^2}\int d^3q\frac{{\bm k}\cdot {\bm q}}{q^2} <\Psi| \delta \hat{\rho}({\bm k}- {\bm q}) \delta \hat{\rho}( {\bm q})|\Psi>,\\
\end{array}
\end{equation}
One can simplify the above equation using the assumption that $\delta\rho({\bm r,t})=\frac{1}{8\pi^3}\int e^{i{\bm k}\cdot{\bm r}}
\delta \tilde{\rho}({\bm k},t)d^3 k$ only weakly varies on the interatomic scale, and hence
three components of the first term in right-hand-side of Eq. (\ref{e10}) can be estimated as:
 $k^2 v_F^2  \delta \tilde{\rho}({\bm k},t)$, $k^3 v_F/k_T\delta \tilde{\rho}({\bm k},t)$  and
 $ k^4 v_F^2/k_T^2\delta \tilde{\rho}({\bm k},t)$, respectively ($1/k_T$ is  Thomas-Fermi  radius\cite{rpa},
 $k_T=\sqrt{\frac{6\pi n_e e^2}{\epsilon_F}}$,
 $\epsilon_F$---the Fermi energy, $v_F$---the Fermi velocity).
Thus the contribution of the second and the third components of the first term can be neglected in comparison to the first component.
 Small and thus negligible is also the last term in right-hand-side of Eq. (\ref{e10}), as it involves a product of two $ \delta \tilde{\rho}$
 (which we assumed small\ $ \delta \tilde{\rho}/n_e << 1$). This approach corresponds to random-phase-approximation (RPA) attitude formulated for
  bulk metal\cite{rpa,pines} (note that $\delta \hat{\rho}(0)=0$ and the coherent RPA contribution of interaction is contained in the second term
   in the right-hand-side of Eq. (\ref{e10})). The last but one term in Eq. (\ref{e10})) can be reduced if to confine  only to linear terms with respect
    to $\delta \tilde{\rho}$ and $\tilde{\varphi}$.
Next, due to spherical symmetry,  $ <\Psi|\sum\limits_{j}e^{-i{\bm k}\cdot {\bm r}_j}\frac{\hbar^2}{m^2}\left( {\bm k}\cdot \nabla_j \right)^2|\Psi>
\simeq \frac{2 k^2}{3m}<\Psi|\sum\limits_{j}e^{-i{\bm k}\cdot {\bm r}_j}\frac{\hbar^2\nabla_j^2}{2m}|\Psi> $.
After the inverse Fourier transform,   Eq. (\ref{e10})  attains  the form:
\begin{equation}
\label{e12}
\begin{array}{l}
\frac{\partial^2 \delta \rho ({\bm r},t) }{\partial t^2}=-\frac{2 }{3m} \nabla^2
<\Psi|\sum\limits_{j}\delta({\bm r}-{\bm r}_j)\frac{\hbar^2\nabla_j^2}{2m}|\Psi>\\
+\frac{\omega_p^2}{4\pi} \nabla \left\{ \Theta(a-r) \nabla \int d^3r_1 \frac{1}{|{\bm r}-{\bm r}_1|} \delta \rho( {\bm r}_1,t)\right\}
+\frac{en_e}{m} \nabla \left\{ \Theta(a-r) \nabla \varphi( {\bm r},t)\right\}.\\
\end{array}
\end{equation}

According to Thomas-Fermi approximation\cite{rpa} the   averaged kinetic energy can be represented as follows:
\begin{equation}
\begin{array}{l}
<\Psi|-\sum\limits_{j}\delta({\bm r}-{\bm r}_j)\frac{\hbar^2\nabla_j^2}{2m}|\Psi>\simeq
\frac{3}{5} (3\pi^2)^{2/3} \frac{\hbar^2}{2m} \rho^{5/3}({\bm r})\\
=\frac{3}{5} (3\pi^2)^{2/3} \frac{\hbar^2}{2m}n_e^{5/3} \Theta(a-r)\left[1+\frac{5}{3}\frac{\delta \rho({\bm r})}{n_e}+...\right].\\
\end{array}
\end{equation}
Taking  into account the above approximation and that $\nabla \Theta(a-r)=-\frac{\bm r}{r}\delta(a-r)= -\frac{\bm r}{r}\lim_{\epsilon\rightarrow 0}\delta(a+\epsilon-r)$
as well as  that $\varphi( {\bm r}, t)=-{\bm r}\cdot {\bm E}(t)$, one can rewrite Eq. (\ref{e12}) in the following manner:
\begin{equation}
\label{e15}
\begin{array}{l}
\frac{\partial^2 \delta \rho ({\bm r},t) }{\partial t^2}=\left[ \frac{2}{3} \frac{\epsilon_F}{m}\nabla^2 \delta \rho( {\bm r},t)-
\omega_p^2 \delta \rho( {\bm r},t)\right]\Theta(a-r)\\
- \frac{2}{3m} \nabla\left\{\left[\frac{3}{5}\epsilon_F n_e+\epsilon_F \delta \rho( {\bm r},t)\right]\frac{\bm r}{r}\delta(a+\epsilon-r)
\right\}\\
-\left[\frac{2}{3} \frac{\epsilon_F}{m}\frac{\bm r}{r}\nabla \delta \rho( {\bm r},t)
+      \frac{\omega_p^2}{4\pi}          \frac{\bm r}{r}\nabla \int d^3r_1 \frac{1}{|{\bm r}-{\bm r}_1|} \delta \rho( {\bm r}_1 ,t)
+\frac{en_e}{m} \frac{\bm r}{r}\cdot{\bm E}(t)\right]
\delta(a+\epsilon-r).\\
\end{array}
\end{equation}
In the above formula  $\omega_p$ is the bulk plasmon frequency, $\omega_p^2=\frac{4\pi n_e e^2}{m}$, and
       $\delta(a+\epsilon-r)   =  \lim_{\epsilon\rightarrow 0}\delta(a+\epsilon-r)$.
The solution of Eq. (\ref{e15}) can be decomposed into two parts with regard  to the domain:
\begin{equation}
 \delta \rho( {\bm r,t})=\left\{
           \begin{array}{l}
             \delta \rho_1( {\bm r,t}), \;for\; r<a,\\
              \delta \rho_2( {\bm r,t}), \;for\; r\geq a,\; ( r\rightarrow a+),\\
          \end{array}
       \right.
       \end{equation}
corresponding to the volume and surface  excitations, respectively. These two parts of local electron density fluctuations
satisfy the equations:
\begin{equation}
\label{e20}
\frac{\partial^2 \delta \rho_1 ({\bm r},t) }{\partial t^2}=\frac{2}{3} \frac{\epsilon_F}{m}\nabla^2 \delta \rho_1( {\bm r},t)-
\omega_p^2 \delta \rho_1( {\bm r},t),
\end{equation}
and
\begin{equation}
\label{e21}
\begin{array}{l}
\frac{\partial^2 \delta \rho_2 ({\bm r},t) }{\partial t^2} =-
\frac{2}{3m} \nabla\left\{\left[\frac{3}{5}\epsilon_F n_e+\epsilon_F \delta \rho_2( {\bm r},t)\right]\frac{\bm r}{r}\delta(a+\epsilon-r)\right\}\\
 -  \left[\frac{2}{3} \frac{\epsilon_F}{m}\frac{\bm r}{r}\nabla \delta \rho_2( {\bm r},t)
+      \frac{\omega_p^2}{4\pi}          \frac{\bm r}{r}\nabla \int d^3r_1 \frac{1}{|{\bm r}-{\bm r}_1|} \left(\delta \rho_1( {\bm r}_1 ,t)
\Theta(a-r_1)\right.\right.\\
\left.\left.+\delta \rho_2( {\bm r}_1 ,t)\Theta(r_1-a)\right)+\frac{en_e}{m} \frac{\bm r}{r}\cdot{\bm E}(t)\right]\delta(a+\epsilon-r).\\
\end{array}
\end{equation}
It is clear from Eq. (\ref{e20}) that the volume plasmons are independent of surface plasmons. However, surface plasmons can
be excited by volume plasmons due to the last term in Eq. (\ref{e21}) (corresponding to 'a surface tail' of volume oscillations),
which expresses a coupling between surface and volume oscillations in the metallic nanosphere within the above semiclassical RPA approach.

In a dielectric medium in which the metallic sphere can be embedded, the electrons on the surface interact with
 forces $\varepsilon$ (dielectric  constant) times weaker in comparison to electrons inside the sphere. To account for it,
  one can substitute Eq.  (\ref{e21}) with the following one (Eq. (\ref{e20}) does not change):
\begin{equation}
\label{e210}
\begin{array}{l}
\frac{\partial^2 \delta \rho_2 ({\bm r},t) }{\partial t^2} =-
\frac{2}{3m} \nabla\left\{\left[\frac{3}{5}\epsilon_F n_e+\epsilon_F \delta \rho_2( {\bm r},t)\right]\frac{\bm r}{r}\delta(a+\epsilon-r)\right\}\\
 -  \left[\frac{2}{3} \frac{\epsilon_F}{m}\frac{\bm r}{r}\nabla \delta \rho_2( {\bm r},t)
+      \frac{\omega_p^2}{4\pi}          \frac{\bm r}{r}\nabla \int d^3r_1 \frac{1}{|{\bm r}-{\bm r}_1|} \left(\delta \rho_1( {\bm r}_1 ,t)
\Theta(a-r_1)\right.\right.\\
\left.\left.+\frac{1}{\varepsilon}\delta \rho_2( {\bm r}_1 ,t)\Theta(r_1-a)\right)+\frac{en_e}{m} \frac{\bm r}{r}\cdot{\bm E}(t)\right]\delta(a+\epsilon-r).\\
\end{array}
\end{equation}

Let us also assume that both volume and surface plasmon oscillations
are damped with the time ratio $\tau_0$ which can be phenomenologically
accounted for via the additional term, $-\frac{2}{\tau_0}\frac{\partial \delta\rho({\bm r},t)}{\partial t}$, to the
 right-hand-side of above equations. They attain the form:
\begin{equation}
\label{e2000}
\frac{\partial^2 \delta \rho_1 ({\bm r},t) }{\partial t^2}+\frac{2}{\tau_0}\frac{\partial \delta\rho_1({\bm r},t)}{\partial t}
=\frac{2}{3} \frac{\epsilon_F}{m}\nabla^2 \delta \rho_1( {\bm r},t)-
\omega_p^2 \delta \rho_1( {\bm r},t),
\end{equation}
and
\begin{equation}
\label{e2100}
\begin{array}{l}
\frac{\partial^2 \delta \rho_2 ({\bm r},t) }{\partial t^2} +\frac{2}{\tau_0}\frac{\partial \delta\rho_2({\bm r},t)}{\partial t}=-
\frac{2}{3m} \nabla\left\{\left[\frac{3}{5}\epsilon_F n_e+\epsilon_F \delta \rho_2( {\bm r},t)\right]\frac{\bm r}{r}\delta(a+\epsilon-r)\right\}\\
 -  \left[\frac{2}{3} \frac{\epsilon_F}{m}\frac{\bm r}{r}\nabla \delta \rho_2( {\bm r},t)
+      \frac{\omega_p^2}{4\pi}          \frac{\bm r}{r}\nabla \int d^3r_1 \frac{1}{|{\bm r}-{\bm r}_1|} \left(\delta \rho_1( {\bm r}_1 ,t)
\Theta(a-r_1)\right.\right.\\
\left.\left.+\frac{1}{\varepsilon}\delta \rho_2( {\bm r}_1, t)\Theta(r_1-a)\right)+\frac{en_e}{m} \frac{\bm r}{r}\cdot{\bm E}(t)\right]\delta(a+\epsilon-r).\\
\end{array}
\end{equation}

From Eqs (\ref{e2000}) and (\ref{e2100}) it is noticeable that the homogeneous electric field does not excite the volume-type plasmons but
 only induces the surface plasmons.

The derived above equations for plasmon excitations in spherical metallic system are in agreement with other similar semiclassical approximations reviewed  e.g.,
in Ref. \onlinecite{kresin}.

\subsection{Solution of RPA equations: volume and surface plasmons frequencies}
Eqs (\ref{e2000}, \ref{e2100}) can be solved upon imposed the boundary and initial
conditions (cf. Appendix \ref{app1}). Let us represent  both parts of the electron fluctuation in the following manner:
\begin{equation}
           \begin{array}{l}
             \delta \rho_1( {\bm r,t})=n_e\left[f_1(r)+F({\bm r}, t)\right], \;for\; r<a,\\
              \delta \rho_2( {\bm r,t})=n_e f_2(r)+\sigma(\Omega,t)\delta(r+\epsilon -a), \;for\; r\geq a,\; ( r\rightarrow a+),\\
          \end{array}
       \end{equation}
and let us choose the convenient initial conditions, $  F({\bm r}, t)|_{t=0}=0, \; \sigma(\Omega,t)|_{t=0}=0$, ($\Omega=(\theta, \psi)$---the spherical angles),
moreover,
$(1+f_1(r))|_{r=a}=f_2(r)|_{r=a}$ (continuity condition), $ F({\bm r}, t)|_{r=a}=0$, $\int\rho({\bm r},t)d^3r=N_e$ (neutrality condition).

We arrive thus with the explicit form of the solutions of   Eqs (\ref{e2000}) and (\ref{e2100}) (cf. Appendix \ref{app1}):
\begin{equation}
\label{fff}
\begin{array}{l}
f_1(r)=-\frac{k_T a +1}{2} e^{-k_T (a-r)} \frac{1-e^{-2k_Tr}}{k_Tr}, \; for \;\;r<a,\\
f_2(r)=\left[k_Ta - \frac{k_Ta+1}{2}\left(1-e^{-2k_Ta}\right)\right]\frac{e^{-k_T(r-a)}}{k_Tr}, \; for\;\; r\geq a,\\
\end{array}
\end{equation}
where $k_T=\sqrt{\frac{6\pi n_e e^2}{\epsilon_F}}=\sqrt{\frac{3\omega_p^2}{v_F^2}}$,
and for time-dependent parts:
\begin{equation}
\label{e2001}
F({\bm r}, t) =\sum\limits_{l=1}^{\infty}\sum\limits_{m=-l}^{l}\sum\limits_{n=1}^{\infty}A_{lmn}j_{l}(k_{nl}r)Y_{lm}(\Omega)sin(\omega'_{nl}t)
e^{-t/\tau_0},
\end{equation}
and
\begin{equation}
\label{e25}
\begin{array}{l}
\sigma(\Omega,t)   = \sum\limits_{l=1}^{\infty}\sum\limits_{m=-l}^{l}Y_{lm}(\Omega)\left[ \frac{B_{lm}}{a^2}sin(\omega'_{0l}t)e^{-t/\tau_0}(1
-\delta_{1l})+ Q_{1m}(t) \delta_{1l}\right]\\
+ \sum\limits_{l=1}^{\infty}\sum\limits_{m=-l}^{l}\sum\limits_{n=1}^{\infty}
A_{lmn}\frac{(l+1)\omega_p^2}{l\omega_p^2-(2l+1)\omega_{nl}^2}Y_{lm}(\Omega)n_e\int\limits_0^a dr_1 \frac{r_1^{l+2}}{a^{l+2}}j_{l}(k_{nl}r_1)
sin(\omega'_{nl}t)e^{-t/\tau_0},\\
\end{array}
\end{equation}
where $j_l(\xi)=\sqrt{\frac{\pi}{2\xi}}I_{l+1/2}(\xi)$ is the spherical Bessel function, $Y_{lm}(\Omega)$ is the spherical function, $\omega_{nl}=
\omega_p\sqrt{1+\frac{x_{nl}^2}{k_T^2a^2}}$ are the frequencies of electron volume free self-oscillations  (volume plasmon frequencies),
$x_{nl}$ are nodes of the Bessel function $j_l(\xi)$, $\omega_{0l}=\omega_p\sqrt{\frac{l}{2l+1}}$ are the frequencies of electron surface free self-oscillations (surface plasmon frequencies), and $k_{nl}=x_{nl}/a$; $\omega'=\sqrt{\omega^2-\frac{1}{\tau_0^2}}$ are the shifted frequencies for all modes due to damping.
  The coefficients $ B_{lm}$ and  $A_{lmn}$ can be  determined by the initial conditions. As we have assumed that $\delta\rho({\bm r}, t=0)=0$,
  we get $ B_{lm}=0$ and  $A_{lmn}=0$, except for  $l=1$ in the former case (of $B_{lm}$), corresponding to response to  homogeneous electric
  field. This mode is described by the function $Q_{1m}(t)$ in the general solution (\ref{e25}).
 The  function $Q_{1m}(t)$ satisfies the equation:
  \begin{equation}
  \label{qqqa}
  \begin{array}{l}
  \frac{\partial^2Q_{1m}(t)}{\partial t^2}+\frac{2}{\tau_0}\frac{\partial Q_{1m}(t)}{\partial t}+\omega_1^2 Q_{1m}(t)\\
   =\sqrt{\frac{4\pi}{3}}\frac{en_e}{m}\left[E_z(t)\delta_{m0}+\sqrt{2}\left(E_x(t)\delta_{m1}
   + E_y(t)\delta_{m-1}\right)\right],\\
   \end{array}
   \end{equation}
   where  $\omega_1=\omega_{01}=\frac{\omega_p}{\sqrt{3\varepsilon}}$ (it is a dipole-type surface plasmon Mie frequency\cite{mie}).
   Only this function contributes the dynamical response to the homogeneous electric field (for the assumed  initial conditions).
From the above  it follows thus that  local electron density (within semiclassical RPA attitude) has the form:
\begin{equation}
\label{e50}
\rho({\bm r},t)=\rho_0(r)+\rho_{1}({\bm r},t),
\end{equation}
with the  RPA  equilibrium electron distribution (correcting the uniform distribution $n_e$):
\begin{equation}
  \rho_0(r)=\left\{
  \begin{array}{l}
  n_e\left[1+f_1(r)\right],\; for\;\; r<a,\\
  n_ef_2(r),\;for\;\; r\geq a, \; r\rightarrow a+\\
  \end{array} \right.
  \end{equation}
and the nonequilibrium part, of surface  plasmon oscillation type:
\begin{equation}
\label{oscyl}
  \rho_{1}({\bm r},t)=\left\{
  \begin{array}{l}
  0,\; for\;\; r<a,\\
\sum\limits_{m=-1}^{1}Q_{1m}(t)Y_{1m}(\Omega)\;for\;\; r\geq a,\; r\rightarrow a+.\\
  \end{array} \right.
  \end{equation}
In general, $ F({\bm r}, t)$ (volume plasmons) and $\sigma(\Omega,t)$ (surface plasmons) contribute to plasmon e-m response. However,
in the case  of homogeneous perturbation, only the surface $l=1$ mode is excited.

For plasmon oscillations given by Eq. (\ref{oscyl}) one can calculate the corresponding dipole,
\begin{equation}
\label{dipol}
{\bm D}(t)= e\int d^3r {\bm r}\rho({\bm r},t)=  \frac{4\pi}{3}e{\bm q}(t)a^3,
\end{equation}
where
$Q_{11}( t)=\sqrt{\frac{8\pi}{3}}q_x( t)$,  $Q_{1-1}( t)=\sqrt{\frac{8\pi}{3}}q_y( t)$,
   $Q_{10}( t)=\sqrt{\frac{4\pi}{3}}q_x(t)$
   and ${\bm q}(t)$ satisfies the equation (cf. Eq. (\ref{qqqa})),
   \begin{equation}
   \label{dipoleq}
   \left[\frac{\partial^2}{\partial t^2}+  \frac{2}{\tau_0}  \frac{\partial}{\partial t} +\omega_1^2\right] {\bm q}(t)=\frac{en_e}{m}
   {\bm E}(t).
   \end{equation}

\section{Lorentz friction for nanosphere plasmons}

The nanosphere plasmons induced by a homogeneous electric field, as described in the above paragraph,
  are themselves a source of the e-m radiation. This radiation takes away the energy of plasmons resulting
  in their damping, which can be described as the Lorentz friction\cite{lan}. This damping was not included in $\tau_0$ in Eq. (\ref{qqqa}).
   The e-m wave emission which causes electron
   friction can be described as the additional electric field\cite{lan},
\begin {equation}
{\bm E}_L= \frac{2}{3\varepsilon v^2}\frac{\partial^3{\bm D}(t)}{\partial t^3},
\end{equation}
where $v=\frac{c}{\sqrt{\varepsilon}}$ is the light velocity in the dielectric medium, and ${\bm D}(t)$ is the dipole of the nanosphere.
According to Eq. (\ref{dipol}) we arrive at the following relation,
\begin{equation}
\label{lor}
{\bm E}_L= \frac{2e}{3\varepsilon v^2}\frac{4\pi}{3}a^3\frac{\partial^3{\bm q}(t)}{\partial t^3}.
\end{equation}
Substituting it into Eq. (\ref{dipoleq}) we get
\begin{equation}
\left[\frac{\partial^2}{\partial t^2}+  \frac{2}{\tau_0}  \frac{\partial}{\partial t} +\omega_1^2\right] {\bm q}(t)=\frac{en_e}{m}
   {\bm E}(t) +\frac{2}{3\omega_1}\left(\frac{\omega_1a}{v}\right)^3\frac{\partial^3{\bm q}(t)}{\partial t^3}.
   \end{equation}
If one assumes the  estimation $ \frac{\partial^3{\bm q}(t)}{\partial t^3}\simeq -\omega_1^2    \frac{\partial{\bm q}(t)}{\partial t}$
 (resulting from perturbative method of solution of the above equation), then one can include the Lorentz friction in a  renormalized damping term:
\begin{equation}
  \left[\frac{\partial^2}{\partial t^2}+  \frac{2}{\tau}  \frac{\partial}{\partial t} +\omega_1^2\right] {\bm q}(t)=\frac{en_e}{m}
   {\bm E}(t) ,
   \end{equation}
   where (cf. Fig. 1),
   \begin{equation}
   \label{tau}
   \frac{1}{\tau}=\frac{1}{\tau_0}+\frac{\omega_1}{3}\left(\frac{\omega_1 a}{v}\right)^3\simeq \frac{v_F}{2\lambda_B}+\frac{Cv_F}{2a}
   +  \frac{\omega_1}{3}\left(\frac{\omega_1 a}{v}\right)^3,
   \end{equation}
 where we used for $\frac{1}{\tau_0}\simeq  \frac{v_F}{2\lambda_B}+\frac{Cv_F}{2a} $
  ($\lambda_B$ is the free path in bulk, $ v_F$ the Fermi velocity, and $C\simeq 1$ is a constant)\cite{atwater,kr}
 which corresponds to inclusion of plasmon damping due to electron scattering on other electrons, on impurities, on phonons  and on nanocrystal boundary.
 The renormalized  damping causes the change in the shift of self-frequency of free surface plasmons,
 $\omega_1'=\sqrt{\omega_1^2-\frac{1}{\tau^2}}$.

 Using Eq. (\ref{tau}) one can determine the radius $a_0$ corresponding to a minimal damping,
 \begin{equation}
 \label{tauo}
 a_0=\frac{\sqrt{3}}{\omega_p}\left(v_Fc^3\sqrt{\varepsilon}/2\right)^{1/4}.
 \label{a0}
 \end{equation}

\subsection{Radiation of surface plasmons on metallic nanosphere in far-field zone}
The Lorentz friction mechanism of energy losses of dipole surface oscillations described above can be also  analyzed equivalently by accounting for e-m emission from oscillating dipoles of plasmons. Let us consider Eq. (\ref{dipoleq}) with irradiation induced damping included into $\tau$ (instead of $\tau_0$).
For $E(t)=E_0[1-\Theta(t)]$ (the rapid switching off  a constant electric field $E_0$) the
solution of Eq. (\ref{dipoleq}) has the form:
\begin{equation}
\label{cosinus}
q(t)=\sqrt{\frac{4\pi}{3}}\frac{en_e}{m\omega_1^2}E_0\left\{\begin{array}{l}
1,\;\; for\;\; t<0,\\
\left[cos(\omega_1 t)+\frac{sin(\omega_1 t)}{\omega_1\tau}\right]e^{-t/\tau},\;\; for\;\;\ t\geq 0,\\
\end{array}\right.
\end{equation}

It is easy to calculate the loss of the total energy of the system, ${\cal{A}}={\cal{E}}(t=0)- {\cal{E}}(t=\infty)$, i.e.,  by taking into account both
kinetic and potential energy of electron system. Only potential interaction energy of oscillating
electrons contributes, and
$ {\cal{E}}(t)=const. +\frac{e^2}{2\varepsilon}a^3q^2(t)$, [the time dependent part of energy is caused by
interaction of  excited electrons,
$\frac{q^2(t)e^2}{2\varepsilon}
\int d^3r_1,d^3r_2\frac{Y_{10}(\Omega_1)\delta(a+\epsilon_1-r_1)     Y_{10}(\Omega_2)\delta(a+\epsilon_2-r_2)}{|{\bm r}_1  -  {\bm r}_2|}$,
with $\epsilon_1,\;    \epsilon_2 \rightarrow 0$, $\epsilon_1>\epsilon_2$]. For $q$ given by Eq. (\ref{cosinus}) we obtain
\begin{equation}
\label{eqA}
{\cal{A}=\cal{E}}(t=0)- {\cal{E}}(t=\infty)= \frac{e^2}{2\varepsilon}a^3\frac{4\pi}{3}\left(\frac{en_eE_0}{m\omega_1^2}\right)^2.
\end{equation}

Radiation of the corresponding dipole,  Eq. (\ref{dipol}), far from the sphere  can be described by  potentials of retarded type, leading to the
formula\cite{lan} for the vector  potential,
${\bm A}({\bm R},t)=\frac{1}{Rc}\frac{\partial {\bm D}\left(t-\frac{R}{v}\right)}{\partial t}$.

 Hence, for far-field radiation of surface plasmon dipole oscillations
 we have
 \begin{equation}
 {\bm B}=rot {\bm A}=-\frac{\sqrt{\varepsilon}}{c^2R}\hat{\bm n}\times\frac{ \partial^2{\bm D}}{\partial t^2},
 \end{equation}
 and
 \begin{equation}
 {\bm E}=\frac{1}{\sqrt{\varepsilon}}{\bm B}\times \hat{\bm n},
 \end{equation}
 corresponding to the planar wave in far-field zone ($\hat{\bm n}={\bm R}/R$),   with the Poyting vector
        ${\bm \Pi}=\frac{v}{4\pi}{\bm E}\times {\bm B}= \frac{\hat{\bm n}}{4 \pi}\frac{\left|\frac{\partial^2
         {\bm D}}{\partial t^2}\right|^2sin^2\Theta}{\varepsilon v^3R^2}$, ($\Theta$
        is the  angle between ${\bm D}$ and ${\bm R}$,  $v=c/\sqrt{\varepsilon}$).
        Next, taking into account that
$ \frac{d \cal{A}}{d t}=\oint {\bm \Pi}\cdot d{\bm s}$, one can find
${\cal{A}} = \int\limits_0^\infty \frac{d{\cal{A}}}{dt} dt
 = \frac{2}{3 \varepsilon v^3}\int\limits_0^\infty \left( \frac{\partial ^2D_z(t-R/v)}{\partial t^2} \right)^2 dt$.
 For $D_z$ given by  Eq. (\ref{dipol}), one can find  the total energy transfer:
 \begin{equation}
 \label{qqq}
 {\cal{A}}=\frac{e^2}{6\varepsilon v^3}\frac{4\pi}{3}a^6\left(\frac{en_eE_0}{m\omega_1^2}\right)^2 \omega_1^4\tau .
 \end{equation}

In this way we estimated the energy loss of plasmon oscillations induced by the  signal $E(t)=E(1-\Theta(t)$),
  which then  irradiated gradually all own energy to the surrounding medium. By comparison of Eqs (\ref{eqA}) and (\ref{qqq}) we find
\begin{equation}
\label{tauu}
 \omega_1\tau=3 \left(\frac{\sqrt{3}c}{a\omega_p}\right)^3.
 \end{equation}

The above calculation of the time ratio $\frac{1}{\tau}$ for oscillation damping due to radiation losses agrees with the
formula for this parameter estimated by Lorentz friction force.

 \subsection{Inclusion of screening effect}

In the above consideration all irradiating electrons in the sphere were treated equivalently. Note that in surface plasmon oscillations take part
all electrons, since it is a translational movement of all collective electrons.  In fact some part of electron irradiation is absorbed
by other electrons in the system, which  reduces outside energy transfer. It can be accounted for in analogy to skin-effect in metals via introducing an effective
radiationally active layer with a depth $h$ close to the sphere surface. Thus the factor $\frac{4\pi}{3}( a^3-(a- h)^3)/\frac{4\pi a^3}{3}$ can be introduced in the formula
(\ref{tauu}) to account for screening skin-effect in the metallic nanosphere, with  $h\sim \frac{1}{\sigma \omega }$, ($\sigma$---conductivity)
as for normal skin-effect\cite{abr}.  Inclusion of screening  results thus in reducing of Lorentz friction and in reducing of  induced by
radiation losses red-shift of resonance, from $a^3$ dependence to
$a^2$, being closer to experimental data in the latter case---cf. Figs 2, 3 and 4. for comparison of $\sim a^3$ and $\sim a^2$ red-shift of resonance.

\section{E-m response of the system of metallic nanospheres}

Let us consider  number $N_s$ of identical metallic nanospheres (with radius $a$) randomly located  in the dielectric medium ($\varepsilon \geq 1$)
of volume $V$. We assume the metal is simple (as considered in the previous sections) and separation between spheres is sufficiently large to
neglect inter-sphere electric interaction. The Hamiltonian of electrons in the system of  $N_s$ spheres  (in 'jellium' model)
has the form: $ \hat{H}_s=\sum\limits_{l=1}^{N_s}\hat{H}_e({\bm r}_l)$, where  ${\bm r}_l$ is the position of $l$-th sphere center, $\hat{H}_e({\bm r}_l)$
is the electron Hamiltonian of the $l$-th sphere. Thus the total electron wave function $\psi_e=\prod\limits_{l=1}^{N_s} \Psi_e^l$,
and $i\hbar\frac{\partial \Psi_e^l}{\partial t}= \hat{H}_e({\bm r}_l) \Psi_e^l $. A density of electrons in the system has the form:
$\rho_s({\bm r}, t)=\sum\limits_{l=1}^{N_s}\rho({\bm r}-{\bm r}_l,t)$, where $ \rho({\bm r}-{\bm r}_l,t)
=<\Psi_e^{l}|\sum\limits_{j} \delta({\bm r}-{\bm r}_l - {\bm r}_j        )|\Psi_e^{l}>$
                                                                 is the contribution to the total electron density  from the
                                                                 $l$th sphere electrons (${\bm r}_j$ is  electron position relative  to sphere center).

The space-time Fourier picture of this electron density has the form:
$\tilde{\rho}_s({\bm k}, \omega)=\frac{1}{(2\pi)^4}\int dt d^3r e^{-i{\bm k}\cdot {\bm r}+i\omega t }\rho_s({\bm r},t)= \tilde{\rho}({\bm k}, \omega)
\sum\limits_{l=1}^{N_s}e^{-i{\bm k}\cdot{\bm r}_l}$. According to the notation (\ref{e50}) we have
$ \tilde{\rho}({\bm k}, \omega)=  \tilde{\rho}_0({\bm k})\delta(\omega)+      \tilde{\rho}_1({\bm k}, \omega)$, with:
\begin{equation}
\begin{array}{l}
\tilde{\rho}_0({\bm k}) =\frac{1}{(2\pi)^3}\int d^3r   \rho_0({\bm r}) e^{-i{\bm k}\cdot {\bm r}},\\
\tilde{\rho}_1({\bm k}, \omega)= \frac{1}{(2\pi)^3}\sum\limits_{l=1}^{\infty}\sum\limits_{m=-l}^{l}
B_{lm}\int d\Omega Y_{lm}(\Omega)e^{-i{\bm k}\cdot{\bm a}}\frac{\delta(\omega+\omega_{0l})-\delta(\omega-\omega_{0l}) }{2i}  \\
+ \frac{1}{(2\pi)^3}\sum\limits_{l=1}^{\infty}\sum\limits_{m=-l}^{l}\sum\limits_{n=1}^{\infty}A_{lm}
\left\{n_e\frac{(l+1)\omega_p^2}{l\omega_p^2-(2l+1)\omega_{nl}^2}\int\limits_{0}^{a} dr_1 \frac{r_1^{2+l}}{a^{2l}}j_l(k_{nl}r_1)\int Y_{ln}(\Omega_1)
e^{-i{\bm k}\cdot{\bm a}} \right.\\
\left. +n_e\int\limits_{0}^{a}dr_1 r_1^2\int d\Omega_1e^{-i{\bm k}\cdot{\bm r_1}} j_l(k_{nl}r_1)Y(\Omega_1)\right\}
\frac{\delta(\omega+\omega_{nl})-\delta(\omega-\omega_{nl}) }{2i},
\end{array}
\end{equation}
here ${\bm a}=a\hat{{\bm r}}_1$, $\hat{{\bm r}}=\frac{{\bm r}}{r}$.
If now one uses the continuity equation, $ \frac{\partial \rho_s}{\partial t}=div {\bm j}_s$, or in the Fourier form,
${\bm k}\cdot \tilde{\bm j}_s({\bm k} , \omega)=\omega \tilde{\rho}_s({\bm k} , \omega)$ (here  $ {\bm j}_s$ is the electron current),
then one can find:  ${\bm k}\cdot \tilde{\bm j}_s({\bm k} , \omega)=\omega \tilde{\rho_1}({\bm k} , \omega)\sum\limits_{l=1}^{N_s}e^{-i {\bm k}\cdot {\bm r}_l}$.

For long wave-length limit ($ka<<1$, which is appropriate for the e-m response of nanospheres; i.e. assuming the 'dipole approximation', when
 only linear in $k$ terms remain)
 we use the following approximations:
$e^{-i{\bm k}\cdot {\bm r}}\simeq 1-i{\bm k}\cdot {\bm r}$ and $  {\bm k}\cdot \tilde{\bm j}({\bm k} , \omega)\simeq  {\bm k}\cdot \tilde{\bm j}(0 , \omega )$.
 Thus one can rewrite the continuity equation in the form:
\begin{equation}
\begin{array}{l}
{\bm k}\cdot \tilde{\bm j}_s(0 , \omega) =N_s\omega\left\{ \sum\limits_{l=1}^{\infty}\sum\limits_{m=-l}^{l}\sum\limits_{n=1}^{\infty}A_{lm}
\left[\frac{n_e}{(2\pi)^3}\int\limits_{0}^{a}dr_1r_1^2j_l(k_{nl}r_1)\int d\Omega_1Y_{lm}(\Omega_1)(-i{\bm k}\cdot{\bm r}_1)\right.\right.\\
 + \frac{n_e}{(2\pi)^3}  \frac{(l+1)\omega_p^2 }{l\omega_p^2-(2l+1)\omega_{nl}^2}\int\limits_{0}^{a}dr_1r_1^{l+2}/a^lj_l(k_{nl}r_1)\int d\Omega_1 Y_{lm}(\Omega_1)
\left. (-i{\bm k}\cdot{\bm a}) \right]      \frac{\delta(\omega+\omega_{nl})-\delta(\omega-\omega_{nl}) }{2i} \\
\left. +\sum\limits_{l=1}^{\infty}\sum\limits_{m=-l}^{l}B_{lm}\frac{1}{(2\pi)^3}\int d\Omega_1 Y_{lm}(\Omega_1) (-i{\bm k}\cdot{\bm a})
                                                     \frac{\delta(\omega+\omega_{0l})-\delta(\omega-\omega_{0l}) }{2i} \right\}\\
                                                     \end{array}
\end{equation}
(as for sufficiently dense nanocomponent system $\sum\limits_{l=1}^{N_s}  e^{-i{\bm k}\cdot {\bm r_l}} \simeq 1$,
and $ \int d\Omega Y_{lm}(\Omega) =0$, for $ l\geq 1$).
Assuming a rapid excitation of all frequencies (${\bm E}(t)= {\bm E}_02\pi \delta(t)$, and thus $\tilde{\bm E}(\omega)={\bm E}_0)$ one can
write from the Ohm law $-e\tilde{{\bm j}}_s(0, \omega)=\sigma(\omega){\bm E}_0$,
 and next, for ${\bm E}_0$ in $z$-th direction,
${\bm k}\cdot{\bm E}_0=cos\Theta ,\;  {\bm k}\cdot{\bm r}_1 =kr_1 \frac{4\pi}{3}\left\{ cos\Theta Y_{10}(\Omega_1)+
\frac{1}{\sqrt{2}}sin(\Theta)[Y_{11}(\Omega_1)+Y_{1-1}(\Omega_1)]\right\}$, and
 $ {\bm k}\cdot{\bm a} =ka \frac{4\pi}{3}\left\{ cos\Theta Y_{10}(\Omega_1')+
\frac{1}{\sqrt{2}}sin(\Theta)[Y_{11}(\Omega_1')+Y_{1-1}(\Omega_1')]\right\}$, where $\Omega_1=(\theta_1, \phi_1)$ and   $\Omega_1'=(\theta_1', \phi_1')$.
Thus one can  obtain:
        \begin{equation}
        \begin{array}{l}
                        \sigma(\omega)E_0 k cos\Theta=
                        \frac{N_{s}|e|\omega k}{2(2\pi)^3}\sqrt{\frac{4\pi}{3}}  \left\{
                        \sum\limits_{n=1}^{\infty}n_e\int\limits_{0}^{a}r_1^3dr_1j_1(k_{1n}r_1)\right.\\
                        \left(1+\frac{2\omega_p^2}{\omega_p^2-3\omega_{n1}^2}\right)\left([A_{10}cos\theta+sin\theta\sqrt{\frac{1}{2}}(A_{11}+A_{1-1})]
                        (\delta(\omega-\omega_{n1})   -  \delta(\omega+\omega_{n1})) \right)\\
                        \left. + a[B_{10}cos\theta +sin\Theta  \sqrt{\frac{1}{2}}(B_{11}+B_{1-1})]
                        \left(\delta(\omega-\omega_{01})   -  \delta(\omega+\omega_{01}) \right)\right\},\\
                     \end{array}
                     \end{equation}
from which $ A_{11}+A_{1-1} = B_{11}+B_{1-1}=0$, while $A_{10}=\mu E_0$ and $B_{10}=\nu n_ea^3E_0$ (the constants $\mu, \nu$ will be  determined
later).   From the above it follows:
\begin{equation}
\begin{array}{l}
\sigma(\omega)=\frac{|e|N_{s}\omega}{2(2\pi)^3}\sqrt{\frac{4\pi}{3}}n_ea^4\left\{
\nu      (\delta(\omega-\omega_{01})   -  \delta(\omega+\omega_{01}))   \right. \\+
\left. \mu\sum\limits_{n=1}^{\infty}\frac{3}{(2k_T^2a^2+3x_{n1}^2)x_{n1}^2}\int\limits_{0}^{x_{n1}}x^3dxj_1(x)
(\delta(\omega-\omega_{n1})   -  \delta(\omega+\omega_{n1})) \right\},
\end{array}
\end{equation}
where $\omega_{n1}^2=\omega_p^2(1+x_{n1}^2/(k_Ta^2))$. Via the formula for $\sigma$ one can now
derive the dielectric response function $\varepsilon(\omega)= \varepsilon'(\omega)  +i   \varepsilon''(\omega)$, with
     $   \varepsilon''(\omega)=\frac{4\pi}{\omega}\sigma(\omega)$ and   $ \varepsilon'(\omega)=\varepsilon+\frac{1}{\pi}
     {\cal{P}}\int\limits_{-\infty}^{+\infty} dx \frac{   \varepsilon''(x)}{x-\omega}$:
     \begin{equation}
     \begin{array}{l}
      \varepsilon''(\omega)=\frac{|e| N_{s} n_e  a^4}{4 \pi^2 }\sqrt{\frac{4\pi}{3}}\left\{\nu  (\delta(\omega-\omega_{01})
       -  \delta(\omega+\omega_{01}))\right.\\
      \left. +\mu \sum\limits_{n=1}^{\infty}\frac{3}{(2k_Ta^2+3x_{n1}^2) x_{n1}^2}\int\limits_{0}^{x_{n1}}dx x^3 j_1(x) (\delta(\omega-\omega_{n1})
       -  \delta(\omega+\omega_{n1}))\right\},\\
   \end{array}
   \end{equation}
   and
     \begin{equation}
     \begin{array}{l}
      \varepsilon'(\omega)=  \varepsilon +\frac{|e| N_{s} n_e  a^4}{2 \pi^2 }\sqrt{\frac{4\pi}{3}}
      \left\{ \nu {\cal P} \frac{\omega_{01}}{\omega_{01}^2-\omega^2} \right.\\
      \left. +\mu \sum\limits_{n=1}^{\infty}\frac{3}{(2k_Ta^2+3x_{n1}^2) x_{n1}^2}\int\limits_{0}^{x_{n1}}dx x^2 j_1(x)
      {\cal P} \frac{\omega_{n1}}{\omega_{n1}^2-\omega^2} \right\}.\\
   \end{array}
   \end{equation}
   One can determine now the constants $\mu, \nu$ using the sum rule: $\int\limits_{0}^{\infty}d\omega \omega \varepsilon''(\omega)=\frac{n2\pi^2e^2}{m}$, where
$n=\frac{N}{V}=\frac{N_s V_0n_e}{V}$, (here, $V$---the volume of the whole system, $V_0$---the volume of the single nanosphere), and the static value
of the dielectric response of the system $\varepsilon(0)=\varepsilon+8\int\limits_{0}^{\infty}d\omega {\cal P}\frac{\sigma(\omega)}{\omega^2}$ (assumed to be known).
 These conditions give:
 \begin{equation}
 \mu=\frac{\omega_p^2 c_0}{u\omega_{01} [\alpha_1-\alpha_2]}\left(1-\frac{\varepsilon(0)-\varepsilon}{3c_0\varepsilon}\right),\;
 \nu= \frac{\omega_p^2 c_0}{u\omega_{01} [\alpha_1-\alpha_2]}\left(\alpha_1 \frac{\varepsilon(0)-\varepsilon}{3c_0\varepsilon}-\alpha_2\right),
 \end{equation}
 where $\alpha_1
 =\sum\limits_{n=1}^{\infty}\frac{3}{(2k_T^2a^2+3x_{n1}^2)x_{n1}^2}\int\limits_{0}^{x_{n1}}dx x^3j_1(x)\frac{\omega_{n1}}{\omega_{01}}$,
$\alpha_2
 =\sum\limits_{n=1}^{\infty}\frac{3}{(2k_T^2a^2+3x_{n1}^2)x_{n1}^2}\int\limits_{0}^{x_{n1}}dx x^3j_1(x)\frac{\omega_{01}}{\omega_{n1}}$,\\
 $c_0=N_s\frac{V_0}{V}$, $ u=\frac{4 |e| n_eN_sa^4}{(2\pi)^3}\sqrt{\frac{4 \pi}{3}}$,.

  By introducing the oscillator strength $f(\omega)=\frac{2\omega\varepsilon''(\omega)}{\pi c_{0}\omega_p^2}$,
 one can express the dipole-type dielectric response of the considered metallically nanomodified system in a more conventional form:
\begin{equation}
\label{e200}
\varepsilon(\omega)=  \varepsilon'(\omega )  +i   \varepsilon''(\omega ) = \varepsilon +c_0 \omega_p^2
\sum\limits_{n=0}^{\infty}\frac{f_n}{2\omega_{n1}}\left[ \frac{1}{\omega_{n1}-\omega-i\epsilon}+  \frac{1}{\omega_{n1}+\omega +i\epsilon}\right],
\end{equation}
where $\epsilon =0+$, $f(\omega)=\sum\limits_{n=0}^{\infty} f_n[\delta(\omega -\omega_{n1})+ \delta(\omega +\omega_{n1}) ]$,
$f_0=\frac{1}{\alpha_1-\alpha_2}\left( \frac{\varepsilon(0)-\varepsilon_0}{3c_0\varepsilon_0}\alpha_1-\alpha_2 \right)$,
$f_n=  \frac{1}{\alpha_1-\alpha_2} \left( 1-   \frac{\varepsilon(0)-\varepsilon_0}{3c_0\varepsilon_0}\right)
\frac{3}{(2k_t^2a^2+3x_{n1}^2)x_{n1}^2}\frac{\omega_{n1}}{\omega_{01}}\int\limits_{0}^{x_{n1}}dx x^3j_1(x)$, $\int\limits_{0}^{\infty}
d\omega f(\omega) =\sum\limits_{n=0}^{\infty}f_n=1$.

We can include  attenuation (also due to irradiation losses),  of dipole excitations ($l=1$), via the
 damping term $\frac{2}{\tau}\frac{\partial \rho_{1}({\bm r},t)}{\partial t}$
in oscillator-type equation for plasmons (which can be
added to left-hand-side of Eq. (\ref{e20}) and assuming that all  modes are damped with the same attenuation time $\tau$).
Thus the time-dependent solution of such modified equation attains the form as given by Eq. (\ref{e2001}) with the factor
$e^{-t/\tau}$, and with the  shifted frequency $\omega'_{n}=\sqrt{\omega_{n}^2-\frac{1}{\tau^2}}$. Similarly to the equation for
the surface plasmons, Eq. (\ref{e21}), can be added (to its left-hand-side) the damping term $\frac{2}{\tau}\frac{\partial \rho_{2}({\bm r},t)}{\partial t}$.
It leads to the factor $e^{-t/\tau}$ for the first part of the solution
(\ref{e25}) (and simultaneously shifted frequency   $\omega'_{l0}=\sqrt{\omega_{l0}^2-\frac{1}{\tau^2}}$), and the second term of
Eq. (\ref{e25}) acquires the additional factor $e^{-t/\tau}$  (and shifted frequency  $\omega'_{n}=\sqrt{\omega_{n}^2-\frac{1}{\tau^2}}$).
The corresponding change in the dipole-type ($l=1$) e-m response function (\ref{e200}) resolves thus to the following expression:
 \begin{equation}
\label{e201}
\varepsilon(\omega)=  \varepsilon'(\omega )  +i   \varepsilon''(\omega ) = \varepsilon +c_0 \omega_p^2
\sum\limits_{n=0}^{\infty}\frac{f_n}{2\omega'_{n1}}\left[ \frac{1}{\omega'_{n1}-\omega-\frac{i}{\tau}}+  \frac{1}{\omega'_{n1}+\omega+\frac{i}{\tau}}\right].
\end{equation}

The above equation can be rewritten as follows:
\begin{equation}
  \varepsilon'(\omega )  = \varepsilon +c_0 \omega_p^2  \sum\limits_{n=0}^{\infty}\frac{f_n}{2\omega'_{n1}}   \left[
\frac{\omega'_{n1}-\omega}{  ( \omega'_{n1}-\omega)^2 +\frac{1}{\tau^2}}    + \frac{\omega'_{n1}+\omega}{  ( \omega'_{n1}+\omega)^2 +\frac{1}{\tau^2}}
\right]
\end{equation}
and
\begin{equation}
  \varepsilon''(\omega )  = c_0 \omega_p^2  \sum\limits_{n=0}^{\infty}\frac{f_n}{2\omega'_{n1}\tau}   \left[
\frac{1}{( \omega^{'}_{n1}-\omega)^2 +\frac{1}{\tau^2}}    - \frac{1}{  ( \omega^{'}_{n1}+\omega)^2 +\frac{1}{\tau^2}}\right].
\end{equation}

\section{Measurement of the dipole surface plasmon frequencies in nanoparticles with variation of their radius}
To determine the plasmon frequencies in metal nanoparticles as a function of the
nanoparticle radius, extinction spectra of colloidal solution of Au and Ag nanoparticles
with radii ranging form 10~nm to 75~nm for Au, and from 10~nm to 40~nm for Ag, respectively,
 were measured. The nanoparticles, prepared as a water colloidal solution with an average
  size distribution not exceeding 8\%
   and an almost constant total mass per ml independent of
   the particle radii, were obtained from British Biocell International. The
    particular data of the Au nanoparticles are listed in Tab. 1.

    \vspace{3mm}

\begin{tabular}{|p{4.5cm}|p{1.2cm}|p{1cm}|p{1cm}|p{1cm}|p{1cm}|p{1cm}|p{1cm}|p{1cm}|}
\multicolumn{4}{c}{Tab. 1. Nanoparticle data for Au colloidal solutions\label{nanoparticle_data}}\\
\hline
nominal nanosphere radius [nm]      & 10    & 15.   & 20    & 25    & 30    & 40    & 50    & 75\\
\hline
average nanosphere radius [nm]      & 10.2  & 15.55 & 20.55 & 24.65 & 29.35 & 39    & 49.45 & 77.15\\
\hline
particle density [10$^9$/ml]		    & 700   & 200   & 90    & 45    & 26    & 11    & 5.6   & 1.7\\
\hline
total volume [10$^{15}$~nm$^3$/ml]		& 3.11  & 3.15  & 3.27  & 2.82  & 2.75  & 2.73  & 2.84  & 3.27\\
\hline
\end{tabular}

\vspace{3mm}

The extinction spectra were measured using a xenon lamp operated at 150~W in combination with a Monospek 1000 monochrometer,
providing monochromatic light at wavelengths from 300~nm to 900~nm. To accommodate for the response of the aqueous environment,
 the cuvette containing the colloidal solution and the lamp spectrum, reference measurements were performed to which the extinction
 spectra are normalized. The extinction coefficient shown in Figs. \ref{Au_extinction_spectra} and \ref{Ag_extinction_spectra} is
  defined as the fraction $T_C/T_{Au,Ag}$, where $T_C$ is the light intensity transmitted through the cuvette containing
   de-ionized water and $T_{Au,Ag}$ light intensity transmitted through the cuvette containing Au or Ag nanoparticle colloidal solutions.

The results are presented in Fig. \ref{Au_extinction_spectra} for Au and in Fig. \ref{Ag_extinction_spectra} for Ag, respectively. The
red-shift of the resonant frequency with growth of nanosphere radius is clearly noticeable. This is accompanied by
the broadening of the attenuation peak and variation of peak height (at the beginning growth and next lowering of the peak height). These
features are collected in the Fig. 2 (bottom) for Au, where the position of the center of extinction peak, its
 half-width and height are plotted versus the
nanosphere radius.

For nanoparticles of gold, silver and copper in the air,
in water and in a colloidal solution, one can find $a_0 \sim 10-14$ nm (cf. Eq. (\ref{tauo}), Fig. 1 and Fig. 4), i.e., the radius of nanosphere
corresponding to minimal damping, which well corresponds to experimental data\cite{ccc,scharte}.
It is a cross-over point for the resonance red-shift versus $a$.  For $a>a_0$ damping increases due to
 Lorentz friction (proportionally to $a^3$, or after inclusion of screening, proportionally to $a^2$) but for $a<a_0$ damping due to electron scattering
 dominates and
  causes  opposite behavior---enhancement of damping with
  lowering radius (proportional to $\frac{1}{a}$, in agreement with experimental observations\cite{atwater,ccc}), which leads to  cross-over
  of
  resonance red-shift  dependence on  $a$.

Surface plasmon oscillations cause attenuation of the incident e-m radiation where the maximum of attenuation is at the resonant frequency\cite{jac}
$\omega_1=\sqrt{\omega_1^2-\frac{1}{\tau^2}}$. This frequency diminishes with growth  of $a$, for $a>a_0$ according to Eq. (\ref{tau}),
which agrees well with the experimental measurements for Au and Ag presented in Figs 2 and 3 and in Tabs 2 and 3 (after inclusion of screening
via skin-effect type correction---cf. Fig. 4).

\vspace{3mm}

\begin{tabular}{|p{6.5cm}|p{1.2cm}|p{1.2cm}|p{1.2cm}|p{1.2cm}|p{1.2cm}|p{1.2cm}|p{1.2cm}|p{1.2cm}|}
\multicolumn{9}{c}{Tab. 2. Resonant frequency for e-m wave attenuation in Au nanospheres \label{Au_resonant_frequencies}}\\
\hline
radius of nanospheres [nm]                    & 10    & 15    & 20    & 25    & 30    & 40    & 50    & 75\\
\hline
$\hbar\omega_1'$ (experiment) [eV]           & 2.371 & 2.362 & 2.357 & 2.340 & 2.316 & 2.248 & 2.172 & 1.895\\
\hline
$\hbar\omega_1'$ (theory, $h=a$) [eV], $n_0= 1.4$   &3.72       &3.716       & 3.71      & 3.69      & 3.66     &  2.41    &    2.37   &XXX \\
\hline
$\hbar\omega_1'$ (theory, $h=a$) [eV], $n_0= 2$     & 2.601      &2.60       & 2.59      & 2.58      & 2.56      &2.38       & 1.64      & XXX\\
\hline
$\hbar\omega_1'$ (theory, $h=6$ nm) [eV], $n_0= 2$   &  2.601     & 2.600      &2.599       & 2.595      & 2.58      &2.55       & 2.47      &1.84\\
\hline
\end{tabular}

($C=2$, $v_F=1.396\cdot 10^6$ m/s,  $\lambda_B=5.3\cdot 10^{-8}$ m,
[cf. Eq. (\ref{tau})], $\omega_p=1.371\cdot 10^{16}$ 1/s, $\omega_1=3.96\cdot10^{15}$ 1/s [for n=2]) XXX---overdamped oscillations;

\vspace{3mm}

\begin{tabular}{|p{6.5cm}|p{1.2cm}|p{1.2cm}|p{1.2cm}|p{1.2cm}|}
\multicolumn{5}{c}{Tab. 3. Resonant frequency for e-m wave attenuation in Ag nanospheres \label{Ag_resonant_frequencies}}\\
\hline
radius of nanospheres [nm]                    & 10    & 20    & 30    & 40\\
\hline
$\hbar\omega_1'$ (experiment) [eV]           & 3.024 & 2.911 & 2.633 & 2.385\\
\hline
$\hbar\omega_1'$ (theory, $h=a$) [eV], $n_0= 1.4$   & 3.71      & 2.70     & 2.66      & 2.41\\
\hline
$\hbar\omega_1'$ (theory, $h=a$) [eV], $n_0= 2$     &  2.61     & 2.60      & 2.56      & 2.382 \\
\hline
$\hbar\omega_1'$ (theory, $h=8$ nm) [eV], $n_0= 2$   & 2.60      & 2.59      & 2.58      &2.55 \\
\hline
\end{tabular}

(C=2, $v_F=1.393 \cdot 10^6$ m/s, $\lambda_B=5.3\cdot 10^{-8}$ m, [cf. Eq. (\ref{tau})], $\omega_p=1.37\cdot 10^{16}$ 1/s, $\omega_1=7.89\cdot 10^{15}$ 1/s [for n=1])
\vspace{3mm}

The observed behavior well corresponds with the formula (\ref{tau}) with the additional skin-effect factor in the last term
$\frac{4\pi}{3}( a^3-(a-h)^3)/\frac{4 \pi}{3} a^3$ (reducing $a^3$ to
$a^2$ radius dependence), which gives  the damping rate $\frac{1}{\tau}$ for surface dipole plasmons versus $a$.
This damping leads to Lorentzian shape of attenuation peak (in response function), $\frac{1/\tau^2}{(\omega-\omega')^2+1/\tau^2}$
with central position at frequency $\omega'=\sqrt{\omega_1^2-1/\tau^2}$. As $1/\tau$ scales as $a^2 $, after inclusion of
screening---cf. Fig. 4,  it is dominating contribution to the value given by Eq. (\ref{tau}) for $a\geq 20$ nm (in agreement with the experiment).
Inclusion of screening (Fig. 4) reduces irradiation-induced damping of plasmons at limiting large values of nanosphere radius and allows to avoid overdamped regime
in this case, which is entered by unscreened $a^3$ damping rate dependence at $a\sim 75$ nm (cf. Tab. 2).
The agreement between the model and the experimental data suggests that for the $a$ dependent red-shift of resonance frequency of surface
 plasmons in metallic nanoparticles of large size ($10<a<75$ nm) responsible are plasmon energy losses caused by Lorentz friction.

\section{Conclusions}

   We have analyzed red-shift of Mie frequency of dipole surface plasmon oscillations in large metallic nanospheres,
    with radius beyond 10 nm. For this region of metallic
   cluster size the dominating channel of plasmon damping starts to be radiation loss due to Lorentz friction. At approximately
     10 nm for nanosphere radius
   the cross-over point of red-shift versus nanosphere radius occurs. For lower radii the $\frac{1}{a}$ rule dominates describing
    scattering Fermi type mechanisms of damping, for higher
   radii the damping due to Lorentz friction, with radius dependence  $\sim a^3$ (or $\sim a^2$ when screening is included),  prevails and quickly
   completely dominates plasmon attenuation.
   The resulting red-shift of
   damped harmonic oscillation well reproduces the experimentally observed resonance positions with respect to metallic sphere size.
   We have measured the resonance  positions via observation of extinction of light
   in water colloidal solutions of Au nanospheres with radii between 10 and 75 nm, and Ag with radii between 10 and 40 nm. The theoretical
   predictions well fit to the experimental behavior, especially if include corrections due to  radiation screening of skin-effect type.
   The resulting
   $a^2$ scaling well reproduces the experimental curves for skin-depth of order of 6 nm (for Au). At the limiting value of metallic nanosphere
    radius ($a>75$ nm) an almost overdamped
   oscillation regime is expected from theoretical analysis, earlier for Ag than for Au, due to  bigger conductivity in Au and thus
    stronger reducing damping   than in Ag. In experiment the corresponding large red-shift in the almost overdamped regime is observed for both Au and Ag.
    Moreover, the first volume mode with
   original energy above $\hbar \omega_p$ (bulk volume plasmon frequency) in the case of an almost overdamped regime is strongly red-shifted
   and emerges in extinction   features as an additional smaller peak
   on the left side of surface plasmon peak for sufficiently large nanospheres, when dipole approximation is not exact.

\begin{acknowledgments}
Supported by the Polish KBN Project No:  N N202 260734 and the FNP Fellowship Start (W. J.), as well as DFG grant SCHA 1576/1-1.

\end{acknowledgments}

\appendix

\section{Analytical solution of plasmon equations for the nanosphere }

\label{app1}

Let us solve first the Eq. (\ref{e2000}), assuming the solution in the form:
\begin{equation}
 \delta\rho_1( {\bm r,t})=n_e\left[f_1(r)+F({\bm r}, t)\right], \;for\; r<a.
 \end{equation}
    Eq. (\ref{e2000}) resolves thus into:
\begin{equation}
\label{ea1}
\begin{array}{l}
\nabla^2 f_1(r)-k_T^2  f_1(r) =0,\\
\frac{\partial^2 F({\bm r},t)}{\partial t^2}+\frac{2}{\tau_0}\frac{\partial F({\bm r},t)}{\partial t}=\frac{v_F^2}{3}\nabla^2 F({\bm r},t) -\omega_p^2  F({\bm r},t).\\
\end{array}
\end{equation}
The solution for function $f_1(r)$ (nonsingular at $r=0$) has thus the form:
\begin{equation}
\label{alpha}
f_1(r)=\alpha \frac{e^{-k_Ta}}{k_Tr}\left(e^{-k_Tr}  -  e^{k_Tr}\right),
\end{equation}
where $\alpha$ is a constant,  $k_T=\sqrt{\frac{6\pi n_e e^2}{\epsilon_F}}=\sqrt{\frac{3\omega_p^2}{v_F^2}}\;$     ($k_T$ is the inverse  Thomas-Fermi radius),
$\omega_p=\sqrt{\frac{4\pi n_e e^2}{m}}\;$ (bulk plasmon frequency).

Since we assumed
 $ F({\bm r},0)=0$, then for function     $ F({\bm r},t)$  the solution can be taken as,
 \begin{equation}
 F({\bm r}, t)= F_{\omega}({\bm r}) sin(\omega' t)e^{-\tau_0t}
 \end{equation}
 where
 $\omega'=\sqrt{\omega^2+ 1/\tau_0^2}$.
$F_{\omega}({\bm r})$ satisfies the equation (Helmholtz equation):
\begin{equation}
\nabla^2  F_{\omega}({\bm r})+k^2 F_{\omega}({\bm r})=0,
\end{equation}
with $k^2=\frac{\omega^2-\omega_p^2}{v_F^2/3}$.
A solution of the above equation, nonsingular at $r=0$, is as follows:
\begin{equation}
 F_{\omega}({\bm r})=Aj_l(kr)Y_{lm}(\Omega),
 \end{equation}
 where $A$ is a constant, $j_l(\xi)=\sqrt{\pi/(2\xi)}I_{l+1/2}(\xi)$ the spherical Bessel function
  [$I_{n}(\xi)$ the Bessel function of the first order],
 and $Y_{lm}(\Omega)$ the spherical function ($ \Omega$ the spherical angle).
Owing to the semiclassical  boundary condition,   $ F({\bm r}, t)|_{r=a}=0$, one has to demand $j_l(ka)=0$, which leads to the discrete values of $k=k_{nl}=x_{nl}/a$,
(where $x_{nl},\;\;n=1,2,3...$, are nodes of $j_{l}$), and next to the discretization of self-frequencies:
\begin{equation}
\omega_{nl}^2=\omega_{p}^2\left(1+\frac{x_{nl}^2}{k_T^2a^2}\right).
\end{equation}

 The general solution for $F({\bm r},t)$ attains thus the form
\begin{equation}
\label{ea10}
F({\bm r},t)=\sum\limits_{l=0}^{\infty}\sum\limits_{m=-l}^{l}\sum\limits_{n=1}^{\infty}
A_{lmn}j_l(k_{nl} r)Y_{lm}(\Omega)sin(\omega'_{nl}t)e^{-\tau_0t}.
\end{equation}

A solution of Eq. (\ref{e2100}) we represent as:
 \begin{equation}
  \delta \rho_2( {\bm r,t})=n_e f_2(r)+\sigma(\Omega,t)\delta(r+\epsilon -a), \;for\; r\geq a,\; ( r\rightarrow a+, i.e. \epsilon \rightarrow 0).
 \end{equation}

The neutrality condition, $\int\rho({\bm r},t)d^3r=N_e$, with
$\delta\rho_2({\bm r},t)=\sigma(\omega, t)\delta(a+\epsilon -r) +n_ef_2(r), (\;\epsilon\rightarrow 0)$, can be rewritten as follows:
$-\int\limits_{0}^{a} dr r^2f_1(r)= \int\limits_{a}^{\infty} dr r^2f_2(r)$, $\int\limits_{0}^{a}d^3rF({\bm r},t )=0$, $\int d\Omega \sigma(\Omega,t)=0$.
Taking into account also  the continuity condition on the   surface, $1+f_1(a)=f_2(a)$, one can obtain:
$f_2(r)=\beta e^{-k_T(r-a)}/(k_Tr)$ and it is possible to fit $\alpha$ (cf. Eq. (\ref{alpha}))  and $\beta$ constants:
$\alpha=\frac{k_Ta +1}{2}$, $\beta=k_Ta -\frac{k_Ta+1}{2}\left(1-e^{-2k_Ta}\right)$, which gives Eqs (\ref{fff}).

 From the condition    $\int\limits_{0}^{a}d^3rF({\bm r},t )=0$  and  from Eq. (\ref{ea10}) it follows  that $A_{00n}=0$,
    (because of $\int d\Omega Y_{lm}(\omega)=4\pi \delta_{l0} \delta_{m0}$).

In order to remove the Dirac delta functions we integrate
both sides of the Eq. (\ref{e2100}) with respect to the radius length  ($\int\limits_{0}^{\infty}r^2dr...$)
 and then we take the limit to the sphere surface, $\epsilon\rightarrow 0$. It results in the following equation
 for surface plasmons:
\begin{equation}
\label{ea15}
\begin{array}{l}
\frac{\partial^2 \sigma(\Omega,t)}{\partial t^2}+\frac{2}{\tau_0}\frac{\partial \sigma(\Omega,t)}{\partial t}=- \sum\limits_{l=0}^{\infty}\sum\limits_{m=-l}^{l}  \omega_{0l}^2
Y_{lm}(\Omega)\int d\Omega_1  \sigma(\Omega_1,t)Y^{*}_{lm}(\Omega_1)\\
+\omega_p^2n_e  \sum\limits_{l=0}^{\infty}\sum\limits_{m=-l}^{l} \sum\limits_{n=1}^{\infty} A_{lmn}
\frac{l+1}{2l+1} Y_{lm}(\Omega)  \int\limits_{0}^{a}dr_1 \frac{r_1^{l+2}}{a^{l+2}}j_l(k_{nl}r_1)sin(\omega_{nl}t),\\
+\frac{en_e}{m}\sqrt{4\pi/3}\left[E_z(t)Y_{10}(\Omega)+\sqrt{2} E_x(t)Y_{11}(\Omega)   + \sqrt{2} E_y(t)Y_{1-1}(\Omega)
\right],
\end{array}
\end{equation}
where $\omega_{0l}^2=\omega_p^2\frac{l}{2l+1}$.
 In derivation of the above equation the following formulae were exploited,
(for $a<r_1$):
    \begin{equation}
                    \frac{\partial}{\partial a}\frac{1}{\sqrt{a^2+r_1^2-2ar_1cos\gamma}}
                    =  \frac{\partial}{\partial a}  \sum\limits_{l=0}^{\infty}\frac{a^l}{r_1^{l+1}}P_l(cos\gamma)= \sum\limits_{l=0}^{\infty}
                    \frac{la^{l-1}}{r_1^{l+1}}P_l(cos\gamma),
                    \end{equation}
where $P_l(cos\gamma)$ is the Legendre polynomial [$P_l(cos\gamma)=\frac{4\pi}{2l+1}\sum\limits_{m=-l}^{l}
Y_{lm}(\Omega)Y^{*}_{lm}(\Omega_1)$], $\gamma$ is an angle between vectors ${\bm a}=a\hat{\bm r}$ and ${\bm r}_1$,
and (for $a>r_1$):
 \begin{equation}
                    \frac{\partial}{\partial a}\frac{1}{\sqrt{a^2+r_1^2-2ar_1cos\gamma}}
                    =  \frac{\partial}{\partial a}  \sum\limits_{l=0}^{\infty}\frac{r_1^l}{a^{l+1}}P_l(cos\gamma)=
                    -\sum\limits_{l=0}^{\infty}\sum\limits_{m=-l}^{l} 4\pi \frac{l+1}{2l+1} \frac{r_1^l}{a^{l+2}}
                    Y_{lm}(\Omega)Y^{*}_{lm}(\Omega_1).
                    \end{equation}

Taking into account the spherical symmetry, one can assume the solution of the Eq. (\ref{ea15}) in the form:
\begin{equation}
 \sigma(\Omega,t) = \sum\limits_{l=0}^{\infty}\sum\limits_{m=-l}^{l}q_{lm}(t)Y_{lm}(\Omega).
 \end{equation}
 From the condition $\int \sigma(\omega t)d\Omega=0$ it follows that $q_{00}=0$. Taking into account the initial condition
 $ \sigma(\omega, 0)=0$ we get (for $l\geq 1$),
 \begin{equation}
 \begin{array}{l}
 q_{lm}(t) =\frac{B_{lm}}{a^2}sin(\omega_{0l}'t)e^{-t/\tau_0}(1-\delta_{l1})+Q_{1m}(t)\delta_{l1}\\
 + \sum\limits_{n=1}^{\infty} A_{lmn}
\frac{(l+1)\omega_p^2}{l\omega_p^2-(2l+1)\omega_{nl}^2} n_e \int\limits_{0}^{a}dr_1
\frac{r_1^{l+2}}{a^{l+2}}j_l(k_{nl}r_1)sin(\omega_{nl}'t)e^{-t/\tau_0},\\
\end{array}
\end{equation}
where $\omega_{0l}'=\sqrt{\omega_{0l}^2-1/\tau_0^2}  $    and  $ Q_{1m}(t)$ satisfies the equation:
\begin{equation}
\frac{\partial^2 Q_{1m}(t}{\partial t^2}+\frac{2}{\tau_0}\frac{\partial Q_{1m}(t)}{\partial t} +\omega_{01}^2Q_{1m}(t)
= \frac{en_e}{m}\sqrt{4\pi/3}\left[E_z(t)\delta_{m0}+\sqrt{2} E_x(t)\delta_{m1}   + \sqrt{2} E_y(t)\delta_{m-1}\right].
\end{equation}
Thus  $ \sigma(\omega, t)$ attains the form:
\begin{equation}
\begin{array}{l}
\sigma(\Omega,t)=\sum\limits_{l=2}^{\infty}\sum\limits_{m=-l}^{l} Y_{lm}(\Omega) \frac{B_{lm}}{a^2}sin(\omega_{0l}'t)e^{-t/\tau_0}+\sum\limits_{m=-1}^{1}
Q_{1m}(t)Y_{1m}(\Omega)\\
 +\sum\limits_{l=1}^{\infty}\sum\limits_{m=-l}^{l}
\sum\limits_{n=1}^{\infty} A_{nlm}
\frac{(l+1)\omega_p^2}{l\omega_p^2-(2l+1)\omega_{nl}^2}Y_{lm}(\Omega)
 n_e \int\limits_{0}^{a}dr_1 \frac{r_1^{l+2}}{a^{l+2}}j_l(k_{nl}r_1)sin(\omega_{nl}'t)e^{-t/\tau_0}.\\
\end{array}
\end{equation}

\newpage

\vspace{1 cm}

\begin{figure}[tb]
\centering
\includegraphics{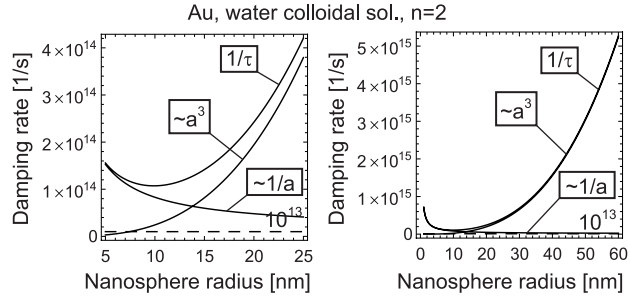}
\caption{\label{Au_data_collection} Damping rate $\frac{1}{\tau}$ of surface plasmons calculated according to the formula (\ref{tau}) versus
nanosphere radius for Au (in water colloidal solution with refraction factor $n=2$ in Eq. (\ref{tau})  $C=2$, $\lambda_B=5.3 \cdot 10^{-8}$ m), dashed line---$10^{13}$ 1/s level }
\end{figure}

\begin{figure}[tb]
\centering
\includegraphics{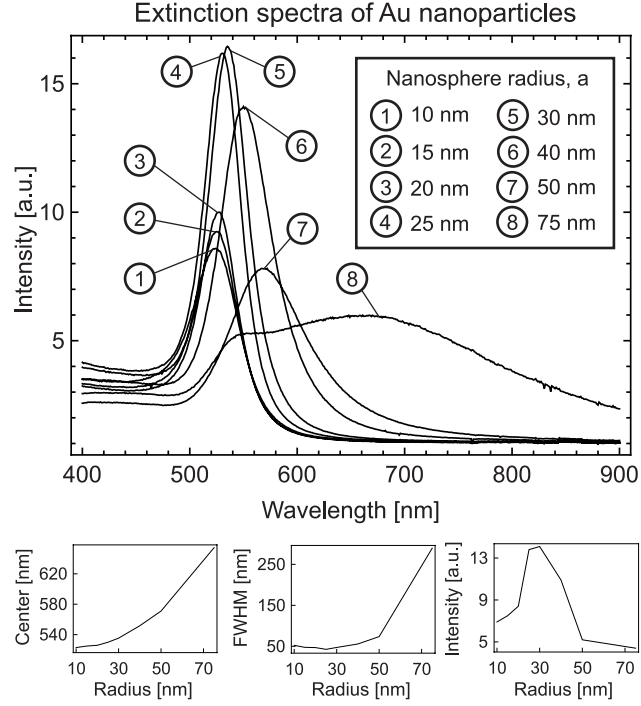}
\caption{\label{Au_extinction_spectra} The results of measurement of light extinction in water colloidal solution of Au nanoparticles
with radii $a$ indicated in the inset; bottom---extracted red-shift of resonance frequency (left), half-width of the attenuation peak (central)
and peak height (right) versus nanosphere radius $a$  }
\end{figure}

\begin{figure}[tb]
\centering
\includegraphics{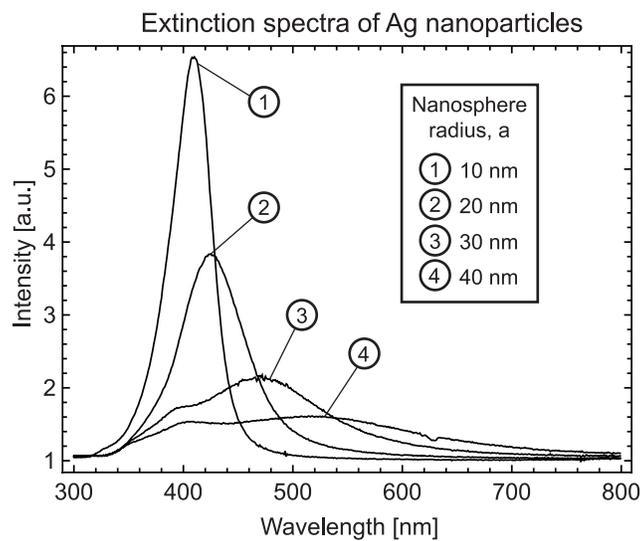}
\caption{\label{Ag_extinction_spectra} The results of measurement of light extinction in water colloidal solution of Ag nanoparticles
with radii $a$ indicated in the inset}
\end{figure}

\begin{figure}[tb]
\centering
\includegraphics{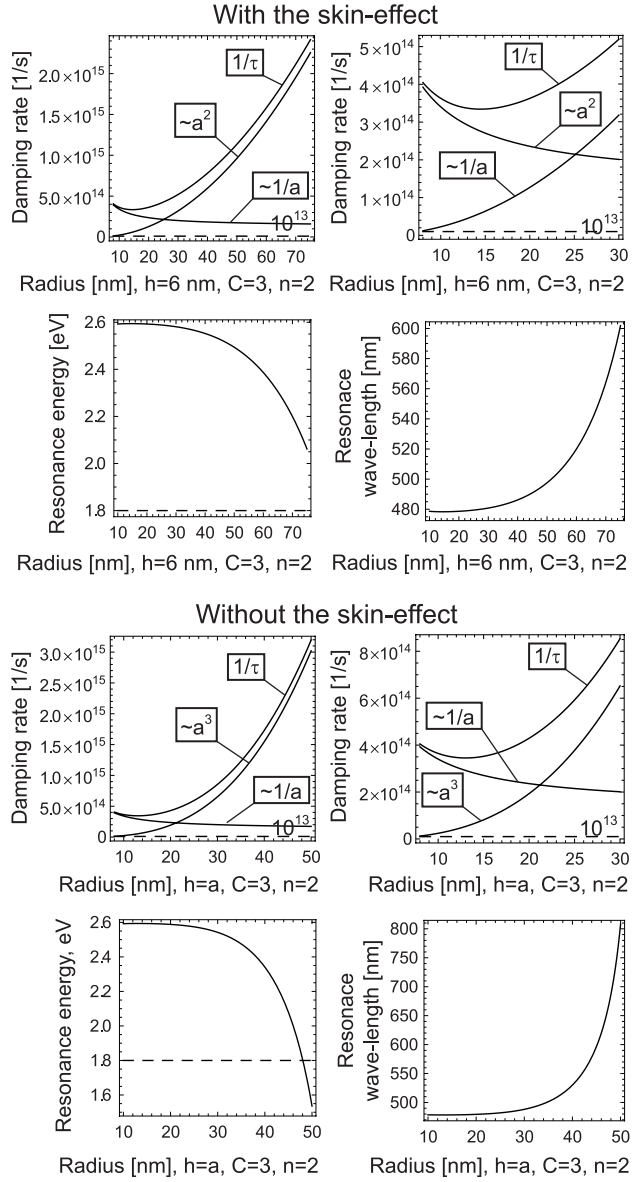}
\caption{\label{Ag_data_collection} Damping rate (Eq. (\ref{tau})) of surface plasmons and corresponding resonance shift with respect to nanosphere radius for Au
in colloidal water solution with inclusion of screening via skin layer with the depth $h=6$ nm (upper) and without screening (lower);
screening (skin-effect) improves
fitting with the experimental data---cf. Fig. 2}
\end{figure}

\end{document}